\documentclass[sigconf, nonacm, balance=false]{acmart}
\usepackage[ruled,vlined]{algorithm2e}
\usepackage{xcolor}
\usepackage{threeparttable}
\AtBeginDocument{%
  \providecommand\BibTeX{{%
    Bib\TeX}}}
\usepackage{booktabs}
\usepackage{multirow}
\usepackage{wrapfig}
\usepackage{hyperref}
\hypersetup{
    colorlinks=true,
    linkcolor=blue,
    urlcolor=blue,
    citecolor=blue
}

\begin{document}

\title{RecRec: Recursive Refinement for Sequential Recommendation}

\author{Pervez Shaik}
\email{shaik.pervez@sony.com}
\affiliation{%
  \institution{Sony Research India}
  \country{India}}

\author{Prosenjit Biswas}
\email{prosenjit.biswas@sony.com}
\affiliation{%
  \institution{Sony Research India}
  \country{India}}

\author{Abhinav Thorat}
\email{Abhinav.Thorat@sony.com}
\affiliation{%
  \institution{Sony Research India}
  \country{India}}

\author{Ravi Kolla}
\email{ravi.kolla@sony.com}
\affiliation{%
  \institution{Sony Research India}
  \country{India}}

\author{Niranjan Pedanekar}
\email{niranjan.pedanekar@sony.com}
\affiliation{%
  \institution{Sony Research India}
  \country{India}}

\renewcommand{\shortauthors}{Shaik et al.}

\begin{abstract}
Sequential recommender systems typically infer user preferences through  single-pass encoding of interaction histories without iterative refinement,  relying on increasingly deep architectures to capture complex patterns.  In this work, we revisit sequential recommendation from a recursive  inference perspective: can user preferences be modeled as a persistent  latent state that is recursively refined? We propose \textbf{RecRec} 
(\underline{Rec}ursive \underline{Rec}ommendation), a lightweight model  that maintains a compact latent state and updates it through a shared recursive module conditioned on interaction evidence. Unlike prior recursive  models, RecRec introduces an evidence-anchored correction mechanism that  stabilizes refinement by grounding each update in the original interaction 
context, preventing semantic drift during deep recursive reasoning.
Experiments on three benchmark datasets under standard evaluation protocols show  that RecRec matches or outperforms state-of-the-art sequential, graph-based, and reasoning-enhanced recommenders while using only 3.9M to 14M parameters. Ablation studies demonstrate that both recursive refinement and the evidence-anchored correction gate contribute significantly to performance, highlighting the effectiveness of recursive latent inference as a scalable alternative to deeper or language-based architectures.
Code is available \href{https://anonymous.4open.science/r/RecRec-6B67/README.md}{\underline{here}}.
\end{abstract}


\maketitle
\section{Introduction}
\label{sec:introduction}
Sequential recommender systems aim to infer user preferences from historical interaction sequences to predict the next relevant item. Existing approaches primarily enhance sequence modeling capacity through Markov models \cite{rendle2010factorizing, he2016fusing}, recurrent networks \cite{liu2016recurrent}, and Transformer-based architectures \cite{qiu2022contrastive,kang2018self}. These methods encode interaction histories in a single forward pass, relying on deeper architectures to capture complex behavioral patterns, lacking mechanisms to iteratively refine or correct preference representations once computed. Consequently, they are limited in their ability to progressively improve estimates or recover from approximation errors. On the other hand, recent reasoning-enhanced approaches based on large and small language models \cite{bismay2025reasoningrec,yao2022react,biswas2025thought} enable multi-step reasoning and inference, but incur substantial computational overhead and rely on token space of the language models.

In this work, we revisit sequential recommendation problem from a recursive inference perspective and ask: \emph{can user preferences be modeled as a persistent latent state that is recursively refined?} We propose \textbf{RecRec} (Recursive Recommendation), a lightweight model that maintains a compact latent state and updates it through a shared recursive module conditioned on interaction evidence, enabling progressive refinement beyond single-pass encoding. A key challenge in recursive refinement is semantic drift, where accumulated approximation errors during recursive updates cause representations to deviate from the original evidence. To address this, we introduce an \emph{evidence-anchored correction mechanism} that grounds each update in the interaction context, ensuring stable refinement. Our approach is inspired by the Tiny Recursive Model (TRM)~\cite{jolicoeur2025less}, but differs by focusing on correction-guided refinement of a persistent latent preference state under direct supervision. 

Our contributions can be summarized as follows:
\begin{itemize}
    \item We formulate sequential recommendation problem as recursive latent state inference problem, modeling user preferences as a persistent latent state refined over multiple steps.
    \item We propose RecRec, a lightweight recursive recommendation model that introduces an evidence-anchored correction mechanism to stabilize recursive updates and prevent semantic drift.
    \item We demonstrate through extensive numerical experiments that RecRec achieves competitive performance against state-of-the-art sequential, graph-based, and reasoning-based recommenders, while maintaining a compact parameter footprint (3.9M--14M) across multiple benchmark datasets and evaluation metrics.

\end{itemize}

\begin{figure*}
\centering
\includegraphics[scale=0.45]{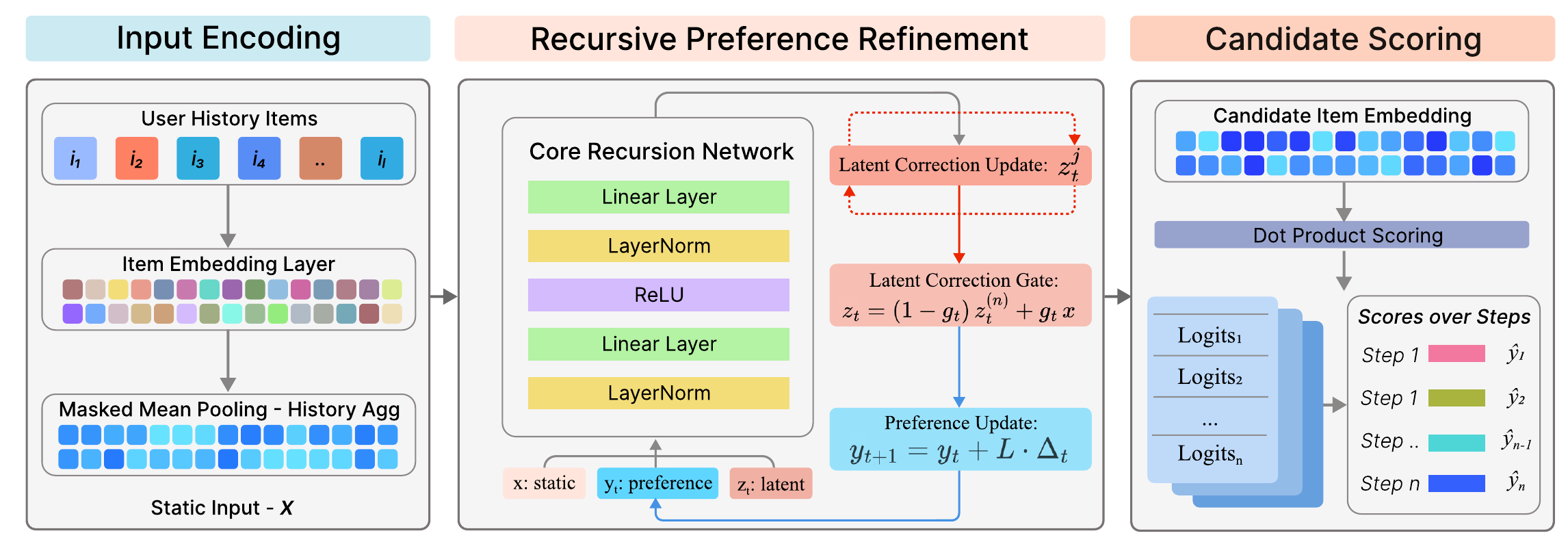}
\caption{RecRec architecture: user history is embedded and aggregated into a static context, then iteratively refined through recursive latent correction and preference updates.}
\label{fig:trm_cf_architecture}
\end{figure*}

\section{Related Work}
\label{sec:literature-survey}

We organize related work into approaches based on sequential modeling, latent reasoning, and language models. \\
\textbf{Sequential Recommendation.}
Early systems modeled user behavior using Markov chains \cite{rendle2010factorizing, he2016fusing} or recurrent neural networks \cite{hidasi2018recurrent, quadrana2017personalizing, liu2016recurrent} to capture short-term dependencies. Transformer-based models \cite{kang2018self, sun2019bert4rec, sawada2025toward} later established strong baselines by utilizing self-attention to model long-range history. However, these methods encode sequences into contextual representations via a single forward pass, lacking the ability to explicitly structure user preference through recursive reasoning. \\
\textbf{Latent Representation and Reasoning.}
To address the lack of structural reasoning in pure sequence models, latent-factor approaches \cite{he2017neural, sedhain2015autorec} learn compact representations from interaction data. This paradigm extends to graph-based \cite{wang2019neural, he2020lightgcn} and knowledge-aware \cite{wang2018ripplenet, wang2019knowledge} methods which propagate embeddings over relational structures. Recent work like ReaRec \cite{tang2025think} attempts to integrate reasoning into this latent space, yet it remains constrained by fixed-depth architectures that do not recursively refine a shared global preference state. \\
\textbf{LLM and SLM-based Recommendation.}
To overcome the limitations of static latent embeddings, recent work explores LLM-based recommenders \cite{bismay2025reasoningrec}, which offer strong reasoning but at high computational cost. SLM approaches \cite{biswas2025thought} provide a lighter alternative with intermediate reasoning, yet rely on explicit language generation rather than compact latent states. Motivated by this, we propose a framework that combines efficient latent modeling with recursive reasoning, detailed in Section \ref{sec:Methodoloy}.

\section{Problem Formulation}
Let $\mathcal{U}$ and $\mathcal{I}$ denote the sets of users and items, respectively. For each user $u \in \mathcal{U}$, we represent their chronologically ordered interaction history as $\mathcal{S}_u = (i_1, i_2, \dots, i_l)$, where $i_j \in \mathcal{I}$ and \textit{l} is the length of the sequence. Each item in the sequence is associated with a pre-trained semantic embedding $\mathbf{e}_{i_j} \in \mathbb{R}^d$ derived from frozen SBERT embeddings of item metadata, providing semantic initialization. We define the set of historical embeddings as $\mathcal{H} = \{\mathbf{e}_{i_1}, \mathbf{e}_{i_2}, \dots, \mathbf{e}_{i_t}\}$. To handle variable sequence lengths during batch processing, we introduce a history mask $\mathcal{M} = (m_1, m_2, \dots, m_t) \in \{0, 1\}^t$, where $m_j=1$ if $i_j$ is a valid interaction and $m_j=0$ otherwise. The objective is to learn a recommendation model that predicts the next item $i_{t+1}$. Following \cite{kang2018self}, to assess a model's performance, items are randomly sampled to form a candidate set $\mathcal{C}_u$ (ground-truth item with sampled negatives), and evaluated under a leave-one-out protocol.

\section{Methodology}
\label{sec:Methodoloy}
In this section, we detail the architecture of \textbf{RecRec}, illustrated in Figure~\ref{fig:trm_cf_architecture}. Our model is designed to recursively update the user’s preference representation while explicitly correcting for semantic drift across steps.

\begin{table*}[t]
\centering
\caption{Performance comparison across three benchmark datasets. RecRec consistently outperforms classical, sequential, and recursive baselines while using fewer parameters.}
\label{tab:main_results}
\resizebox{\textwidth}{!}{%
\begin{tabular}{@{}l|c|ccc|ccc|ccc|c|ccc|ccc|ccc|c|ccc|ccc|ccc@{}}
\toprule
\textbf{Model} & \multicolumn{10}{c|}{\textbf{Luxury Beauty}} & \multicolumn{10}{c|}{\textbf{Video Games}} & \multicolumn{10}{c}{\textbf{Steam Games}} \\
\cmidrule(l){2-31}
 & \textbf{Param}
 & \multicolumn{3}{c|}{\textbf{HR@k}}
 & \multicolumn{3}{c|}{\textbf{NDCG@k}}
 & \multicolumn{3}{c|}{\textbf{Prec@k}}
 & \textbf{Param}
 & \multicolumn{3}{c|}{\textbf{HR@k}}
 & \multicolumn{3}{c|}{\textbf{NDCG@k}}
 & \multicolumn{3}{c|}{\textbf{Prec@k}}
 & \textbf{Param}
 & \multicolumn{3}{c|}{\textbf{HR@k}}
 & \multicolumn{3}{c|}{\textbf{NDCG@k}}
 & \multicolumn{3}{c}{\textbf{Prec@k}} \\
 & & 1 & 5 & 10 & 1 & 5 & 10 & 1 & 5 & 10
   & & 1 & 5 & 10 & 1 & 5 & 10 & 1 & 5 & 10
   & & 1 & 5 & 10 & 1 & 5 & 10 & 1 & 5 & 10 \\
\midrule
LightGCN  & 7M    & 0.37 & 0.49 & 0.55 & 0.37 & 0.41 & 0.47 & 0.37 & 0.11 & \underline{0.06} & 37M   & 0.30 & 0.57 & 0.67 & 0.30 & 0.44 & 0.48 & 0.30 & \underline{0.11} & \underline{0.06} & 30M   & 0.29 & 0.72 & 0.83 & 0.29 & 0.51 & 0.55 & 0.29 & 0.14 & \underline{0.08} \\
NGCF      & 7.9M  & 0.35 & 0.51 & 0.60 & 0.35 & 0.43 & 0.46 & 0.35 & 0.10 & \underline{0.06} & 38M   & 0.25 & 0.51 & 0.63 & 0.25 & 0.39 & 0.42 & 0.25 & 0.10 & \underline{0.06} & 30.7M & 0.41 & 0.80 & 0.87 & 0.41 & \underline{0.62} & \underline{0.68} & 0.41 & 0.16 & \textbf{0.09} \\
SasRec    & 6.7M  & 0.30 & 0.41 & 0.50 & 0.30 & 0.35 & 0.38 & 0.30 & 0.08 & 0.05 & 17.5M & 0.27 & 0.51 & 0.62 & 0.27 & 0.39 & 0.43 & 0.27 & 0.10 & \underline{0.06} & 7.3M  & 0.43 & 0.73 & 0.85 & 0.43 & 0.59 & 0.62 & 0.43 & 0.14 & \underline{0.08} \\
UniRec    & 8M    & 0.36 & 0.47 & 0.52 & 0.36 & 0.41 & 0.43 & 0.36 & 0.06 & 0.04 & 18.4M & 0.31 & 0.55 & 0.66 & 0.31 & 0.43 & 0.48 & 0.31 & 0.08 & 0.05 & 10M   & 0.41 & 0.73 & 0.82 & 0.41 & 0.58 & 0.61 & 0.41 & 0.14 & \underline{0.08} \\
ReaRec    & 5M    & 0.41 & 0.54 & \underline{0.61} & 0.41 & 0.48 & \underline{0.50} & 0.41 & \textbf{0.12} & \textbf{0.07} & 16.4M & 0.33 & \textbf{0.61} & 0.69 & 0.33 & \underline{0.48} & \underline{0.51} & 0.33 & \textbf{0.12} & \textbf{0.07} & 6.7M  & 0.36 & 0.69 & 0.80 & 0.36 & 0.53 & 0.57 & 0.36 & 0.14 & \underline{0.08} \\
\midrule
{\textbf{RecRec - SBERT}}  & {\textbf{3.9M}}  & {\textbf{0.43}} & {\textbf{0.59}} & {\textbf{0.66}} & {\textbf{0.43}} & {\textbf{0.51}} & {\textbf{0.53}} & {\textbf{0.43}} & {\textbf{0.12}} & {\underline{0.06}} & {\textbf{13.8M}} & {\textbf{0.36}} & {\textbf{0.61}} & {\textbf{0.71}} & {\textbf{0.36}} & {\textbf{0.49}} & {\textbf{0.52}} & {\textbf{0.36}} & {\textbf{0.12}} & {\textbf{0.07}} & {\textbf{5.2M}} & {\underline{0.59}} & {\textbf{0.87}} & {\textbf{0.93}} & {\underline{0.59}} & {\textbf{0.74}} & {\textbf{0.76}} & {\underline{0.59}} & {\underline{0.17}} & {\textbf{0.09}} \\
{\textbf{RecRec - Random}} & {\textbf{3.9M}}  & {\underline{0.42}} & {\underline{0.57}} & {\textbf{0.66}} & {\underline{0.42}} & {\underline{0.50}} & {\textbf{0.53}} & {\underline{0.42}} & {\underline{0.11}} & {\underline{0.06}} & {\textbf{13.8M}} & {\underline{0.35}} & {\underline{0.59}} & {\underline{0.70}} & {\underline{0.35}} & {\underline{0.48}} & {\underline{0.51}} & {\underline{0.35}} & {\textbf{0.12}} & {\textbf{0.07}} & {\textbf{5.2M}} & {\textbf{0.60}} & {\underline{0.86}} & {\underline{0.92}} & {\textbf{0.60}} & {\textbf{0.74}} & {\textbf{0.76}} & {\textbf{0.60}} & {\textbf{0.18}} & {\textbf{0.09}} \\
\midrule
\end{tabular}%
}
\end{table*}
\subsection{Recursive Refinement with Self-Correction}
RecRec performs iterative preference refinement over $T$ steps using a shared recursive block. Instead of a single hidden representation, the model maintains three states: a static context $x$, a preference state $y_t$, and a latent state $z_t$. \\
\textbf{Context Initialization:} The context $x$ is obtained via masked mean pooling to frozen SBERT embeddings of the interaction history $\mathcal{H},$ as follows: $x = \sum (\text{Embed}(\mathcal{H}) \odot \mathcal{M})/\sum \mathcal{M},$ where $\odot$ denotes the inner product operator.
We initialize $y_0 \leftarrow x$ and $z_0 \leftarrow \mathbf{0}$. \\
\textbf{Refinement Loop:} At each outer step $t \in \{0, \dots, T-1\}$, the model performs $n$ inner recursive updates to stabilize the latent representation $z_t.$ During this process, the preference state $y_t$ remains fixed, while the latent state $z$ is iteratively updated using a shared non-linear transformation $f_{\phi}$. Specifically, for each inner step $j \in \{1, \dots, n\}$:
\begin{equation}
\label{eq:inner_refine}
z^{(j)}_t = f_{\phi}([x \,\|\, y_t \,\|\, z^{(j-1)}_t]),
\end{equation}
where $[\cdot\|\cdot]$ denotes concatenation operator and $z^{(0)}_t$ is inherited from the previous outer step. This inner recursion progressively refines $z$ conditioned on both the static evidence $x$ and the current preference estimate $y_t$, allowing the model to iteratively reconcile contextual information before updating the preference state. By applying the same transformation across steps, the model traces a structured trajectory in latent space rather than relying on a single-pass transformation. \\
\textbf{Self-Correction Mechanism:} A key challenge in recursive refinement is \emph{semantic drift}, where repeated application of a shared nonlinear transformation causes the latent state to deviate from the original evidence. To mitigate this, we introduce a \textit{correction gate}~$g_t$ that adaptively controls the influence of the original context $x$ during refinement, as given below:
\begin{equation}
\label{eq_gate}
g_t = \sigma(W_t [x \,\|\, y_t]),
\end{equation}
where $W_t$ is a learnable parameter matrix of gate correction mechanism and $\sigma$ denotes the sigmoid function. Given the final inner-loop state $z_t^{(n)}$, we compute a corrected latent state as follows:
${z}_{t} = (1 - g_t) \, z_t^{(n)} + g_t \, x,$ which interpolates between the refined representation and the original evidence. Note that $g_t$ can be interpreted as a data-dependent trust coefficient that regulates the contribution of recursive updates. This mechanism explicitly anchors the recursive updates to the observed interaction history, preventing error accumulation and ensuring stable refinement across steps. Unlike standard deep models where such alignment is implicit, the correction gate provides a direct and adaptive pathway for evidence injection during inference. \\
\textbf{Preference Refinement:}
The preference state, $y_t,$ is updated using the corrected latent representation ${z}_t$ as follows:
\begin{equation}
\label{eq:intent_update}
{y}_{t+1} = y_t + L \cdot \underbrace{\tanh\!\left(f_\phi\!\left([x \,\|\, y_t \,\|\, \tilde{z}_t]\right)\right)}_{\Delta_t},
\end{equation}
where $L$ is a scaling factor controlling the update magnitude. Note that, ${y}_{t+1}$ is the updated state for the next iteration. This residual formulation produces an incremental update $\Delta_t$ conditioned on both the current preference state and the corrected latent representation. By iteratively applying small, controlled updates, the model refines user preferences over multiple steps while maintaining stability.
\subsection{Deep Supervision and Learning Objective}
\begin{figure}
\includegraphics[scale=0.35]{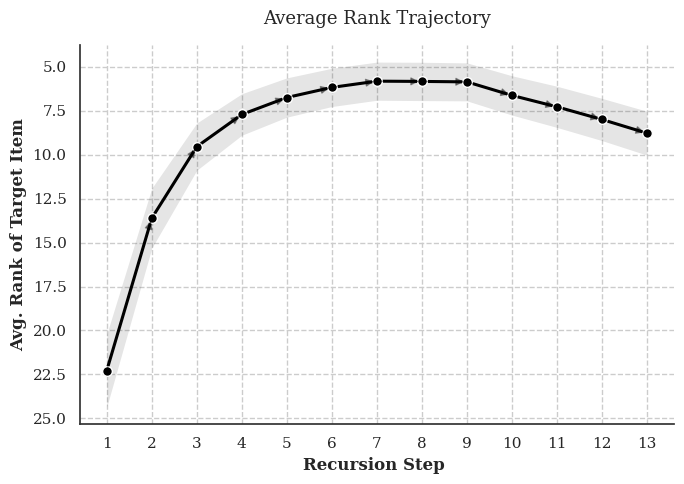}
\caption{Step-wise average rank improvement}
\label{fig:rank_improvement}
\end{figure}
To effectively train the shared parameters $\phi$ in Eq.~\ref{eq:inner_refine} and $W_t$ in Eq.~\ref{eq_gate} while mitigating vanishing gradients across recursive steps, we employ deep supervision. Starting from an initial state $y_0 = x$, the model performs $T$ refinement steps. At each step $t \in \{0, \dots, T-1\}$, an updated representation ${y}_{t+1}$ is computed and used for supervision against the target item $i_{\text{target}}$ using the cross-entropy loss as follows:
\begin{equation}
\mathcal{L}_{CE}({y}_{t+1}, i_{\text{target}}) = 
- \log \left(
\frac{
\exp({y}_{t+1} \cdot \mathbf{e}_{\text{target}}^\top / \tau)
}{
\sum_{j \in \mathcal{C}} 
\exp({y}_{t+1} \cdot \mathbf{e}_j^\top / \tau)
}
\right),
\end{equation}
where $\mathbf{e}_j$ denotes the embedding of candidate item $j$, $\mathcal{C}$ is the candidate set consisting of the ground-truth item and sampled negatives, $\tau$ is the softmax temperature. The total loss is defined as:
$
\mathcal{L}_{\text{total}} = \frac{1}{T} \sum_{t=0}^{T-1} 
\mathcal{L}_{CE}({y}_{t+1}, i_{\text{target}}).
$
This step-wise supervision encourages the model to produce progressively refined and predictive representations at each recursion step, rather than relying solely on the final output. Gradients are backpropagated through both the inner recursion (Eq.~\ref{eq:inner_refine}) and the outer refinement updates (Eq.~\ref{eq:intent_update}), enabling stable optimization of the shared recursive core. This contrasts with standard sequential recommenders that apply supervision only at the final prediction layer.
\begin{table}[ht]
\centering
\scriptsize
\setlength{\tabcolsep}{4pt}
\caption{Ablation study on Steam dataset.}
\label{tab:ablation_steam}
\begin{tabular}{@{}l|ccc|ccc|ccc@{}}
\toprule
\textbf{Configuration} 
& \multicolumn{3}{c|}{\textbf{HR@k}} 
& \multicolumn{3}{c|}{\textbf{NDCG@k}} 
& \multicolumn{3}{c}{\textbf{Prec@k}} \\
\cmidrule(lr){2-4} \cmidrule(lr){5-7} \cmidrule(lr){8-10}
& 1 & 5 & 10 & 1 & 5 & 10 & 1 & 5 & 10 \\
\midrule
No correction gate (G) & 0.46 & 0.75 & 0.83 & 0.46 & 0.62 & 0.64 & 0.46 & 0.15 & 0.08 \\
No inner recursion (R) & 0.53 & 0.80 & 0.87 & 0.53 & 0.68 & 0.71 & 0.53 & 0.16 & 0.08 \\
No latent $Z$          & 0.48 & 0.77 & 0.84 & 0.48 & 0.63 & 0.66 & 0.48 & 0.15 & 0.08 \\
\midrule
\textbf{Full (Z + R + G)} 
& \textbf{0.59} & \textbf{0.87} & \textbf{0.93} 
& \textbf{0.59} & \textbf{0.74} & \textbf{0.76} 
& \textbf{0.59} & \textbf{0.17} & \textbf{0.09} \\
\bottomrule
\end{tabular}
\end{table}

\section{Experiments and Results}
\label{sec:Results}
\begin{table*}
\centering
\scriptsize
\setlength{\tabcolsep}{3pt}
\caption{Performance comparison with LLM baselines across three benchmark datasets.}
\label{llm_baselines}
\resizebox{\textwidth}{!}{%
\begin{tabular}{@{}l|c|ccc|ccc|ccc|c|ccc|ccc|ccc|c|ccc|ccc|ccc@{}}
\toprule
\textbf{Model} & \multicolumn{10}{c|}{\textbf{Luxury Beauty}} & \multicolumn{10}{c|}{\textbf{Video Games}} & \multicolumn{10}{c}{\textbf{Steam Games}} \\
\cmidrule(l){2-31}
 & \textbf{Param}
 & \multicolumn{3}{c|}{\textbf{HR@k}}
 & \multicolumn{3}{c|}{\textbf{NDCG@k}}
 & \multicolumn{3}{c|}{\textbf{Prec@k}}
 & \textbf{Param}
 & \multicolumn{3}{c|}{\textbf{HR@k}}
 & \multicolumn{3}{c|}{\textbf{NDCG@k}}
 & \multicolumn{3}{c|}{\textbf{Prec@k}}
 & \textbf{Param}
 & \multicolumn{3}{c|}{\textbf{HR@k}}
 & \multicolumn{3}{c|}{\textbf{NDCG@k}}
 & \multicolumn{3}{c}{\textbf{Prec@k}} \\
 & & 1 & 5 & 10 & 1 & 5 & 10 & 1 & 5 & 10
   & & 1 & 5 & 10 & 1 & 5 & 10 & 1 & 5 & 10
   & & 1 & 5 & 10 & 1 & 5 & 10 & 1 & 5 & 10 \\
\midrule

Allmrec & 7B 
& \underline{0.41} & \underline{0.73} & \underline{0.80} & \underline{0.41} & \underline{0.61} & \underline{0.64} & \underline{0.41} & 0.15 & 0.07
& 7B & \underline{0.47} & \underline{0.77} & 0.82 & \underline{0.47} & 0.55 & 0.58 & \underline{0.47} & \underline{0.17} & 0.08
& 7B & 0.51 & 0.83 & \underline{0.89} & 0.51 & 0.67 & 0.69 & 0.51 & \textbf{0.23} & \textbf{0.12} \\

Tallrec & 7B 
& 0.42 & \underline{0.73} & 0.79 & 0.42 & 0.57 & 0.60 & 0.42 & 0.13 & 0.06
& 7B & 0.42 & 0.75 & \underline{0.83} & 0.42 & \underline{0.60} & \underline{0.62} & 0.42 & \textbf{0.18} & \underline{0.09}
& 7B & \underline{0.59} & \underline{0.87} & 0.92 & \underline{0.59} & \underline{0.68} & \underline{0.72} & \underline{0.59} & \underline{0.20} & \textbf{0.12} \\

Gpt4rec & 117M 
& 0.37 & 0.69 & 0.75 & 0.37 & 0.54 & 0.57 & 0.37 & \textbf{0.17} & \textbf{0.09}
& 117M & 0.39 & 0.71 & 0.77 & 0.39 & 0.53 & 0.55 & 0.39 & \underline{0.17} & 0.08
& 117M & 0.39 & 0.75 & 0.80 & 0.39 & 0.59 & 0.63 & 0.39 & \underline{0.20} & \underline{0.11} \\

RecRec & \textbf{3.9M} 
& \textbf{0.49} & \textbf{0.80} & \textbf{0.88} & \textbf{0.49} & \textbf{0.65} & \textbf{0.68} & \textbf{0.49} & \underline{0.16} & \underline{0.08}
& \textbf{13.8M} & \textbf{0.53} & \textbf{0.83} & \textbf{0.91} & \textbf{0.53} & \textbf{0.70} & \textbf{0.72} & \textbf{0.53} & 0.16 & \underline{0.09}
& \textbf{5.2M} & \textbf{0.70} & \textbf{0.90} & \textbf{0.95} & \textbf{0.70} & \textbf{0.80} & \textbf{0.82} & \textbf{0.70} & 0.19 & 0.09 \\

\bottomrule
\end{tabular}%
}
\end{table*}
\subsection{Datasets}
We evaluate RecRec on three publicly available datasets~\cite{hou2024bridging} from different domains: \textit{Luxury Beauty} (11,690 users, 6,534 items, 71,898 interactions), \textit{Video Games} (64,039 users, 33,611 items, 508,508 interactions), and \textit{Steam Games} (67,610 users, 10,978 items, 5,023,170 interactions). All datasets were preprocessed to remove users and items with fewer than five interactions prior to SBERT embedding extraction. These datasets cover diverse domains and interaction densities, enabling evaluation across varying sparsity and sequence characteristics. Implementation details can be found in \href{https://anonymous.4open.science/r/RecRec-6B67/Appendix.pdf}{\underline{Appendix}}.
\subsection{Comparative Performance Evaluation}

We compare RecRec against classical, sequential, recursive, and LLM-based recommenders. Across all datasets, RecRec achieves consistent gains while maintaining a substantially smaller parameter footprint.\\
\textbf{Comparison with Sequential and Graph-based Recommenders.}
As shown in Table~\ref{tab:main_results}, RecRec consistently outperforms established baselines, including graph-based models (LightGCN~\cite{he2020lightgcn}, NGCF~\cite{wang2019neural}) and sequential architectures (SASRec~\cite{kang2018self}, UniRec~\cite{liu2024unirec}). Following~\cite{kang2018self}, we adopt an interaction history length of 50 and a candidate set size of 100 for all models in Table~\ref{tab:main_results}. Despite this more challenging setting, RecRec outperforms all baselines, while maintaining substantially smaller parameter footprint (e.g., only 3.9M parameters on Luxury Beauty). The recursion-based baseline ReaRec~\cite{tang2025think} provides the strongest competing performance; however, RecRec preserves a consistent advantage, particularly on early-ranking metrics such as HR@1 and NDCG@1.
Importantly, this improvement is not driven by pre-trained semantic initialization. The RecRec(Random) variant, which replaces SBERT embeddings with random uniform initialization, achieves comparable performance  ($\pm2\%$), indicating that gains primarily arise from recursive refinement and latent stabilization rather than external semantic priors.\\
\textbf{Comparison with LLM-based Baselines.} We additionally compare RecRec against three representative LLM-based recommenders spanning key design paradigms: ALLMRec~\cite{kim2024large}, which aligns LLM token representations with collaborative filtering objectives; TallRec~\cite{bao2023tallrec}, a state-of-the-art large-scale (7B) sequential recommender; and GPT4Rec~\cite{li2023gpt4rec}, a parameter-efficient variant that adapts pretrained LLMs for recommendation. To ensure a standardized comparison, we adopted the configuration utilized for all three baselines, setting the interaction history length to 10 and the candidate set size to 20 following \cite{kim2024large}. Furthermore, we utilized a point-wise scoring evaluation setting, consistent with the methodology in TallRec~\cite{bao2023tallrec}, where the model calculates individual likelihood scores for each candidate in the set. This differs from the setting in Table~\ref{tab:main_results} (history = 50, candidates = 100), leading to uniformly higher absolute HR@1 values across all models. As shown in Table~\ref{llm_baselines}, RecRec consistently outperforms these LLM-based approaches across multiple datasets. \\
\textbf{Effect of Recursion Steps ($T$).} 
Our analysis on the Steam dataset shows that at $T=1$, RecRec reduces to a single-pass architecture (HR@1: 0.49). Increasing $T$ leads to consistent performance gains, with HR@1 peaking at $T=7$ (HR@1: 0.59). Beyond this point, performance degrades (e.g., HR@1 drops to 0.55 at $T=13$), indicating over-refinement. As shown in Fig.~\ref{fig:rank_improvement}, the average rank improves up to $T \approx 6\text{--}7$ before plateauing and slightly degrading. A consistent trend is observed in HR@1, which peaks at $T=7$ and declines thereafter. This suggests that while early recursion effectively refines user intent, deeper recursion yields diminishing returns, potentially due to error accumulation in repeated nonlinear updates.\\
\textbf{Parametric Efficiency.}
RecRec exhibits superior parameter efficiency, outperforming the next-best baseline, ReaRec~\cite{tang2025think}, with 22\% fewer parameters. Remarkably, RecRec surpasses LLM-based models like ALLMRec and GPT4Rec despite a 99\% smaller footprint (3.9M vs. 7B). \\
\textbf{Effect of Model Depth.}
We vary the depth of the core MLP, $f_{\phi}$, given in Equation~\ref{eq:inner_refine}, from 2 to 10 layers on the Steam dataset. Performance is highest at 5 layers (HR@1: 0.61, HR@10: 0.94) with 6.2M parameters, improving from HR@1: 0.60 at 2 layers. Increasing the depth to 10 layers (21.1M parameters) degrades performance (HR@1: 0.55), indicating overfitting to dataset-specific patterns rather than improved generalization. Overall, this suggests that recursive refinement provides a more parameter-efficient alternative to increasing network depth, as deeper models tend to overfit rather than enhance preference estimation.

\subsection{Ablations}
Ablation results for the core components of RecRec, namely the correction gate, inner recursion, and latent state, are shown in Table~\ref{tab:ablation_steam}. Due to space constraints, we present results only on the Steam Games dataset; corresponding results for the other datasets are provided in the Appendix. First, removing the \textit{correction gate} yields the largest degradation across all metrics, compared to removing either the inner recursion or the latent state, underscoring its critical role in stabilizing recursive updates and mitigating error accumulation. Next, removing the latent state, $z$, which serves as an internal reasoning representation, forces memory-less updates and leads to a substantial performance drop, as the model can no longer maintain a persistent summary of interaction history across recursion steps. Finally, excluding the \textit{inner recursion} mechanism consistently degrades performance, highlighting the importance of iterative refinement within the weight-shared core for learning a stable latent state prior to the final update to $y.$

\section{Conclusion}
\label{sec:Conclusion}

We presented RecRec, a lightweight sequential recommender that models user preferences as a recursively refined latent state. RecRec achieves competitive performance through iterative refinement with only 3.9M–14M parameters, consistently matching or outperforming strong baselines across multiple benchmarks. Ablation studies show that gains arise from structured recursion, correction-gated stabilization, and persistent latent state modeling rather than increased capacity or external semantic priors. Our analysis further reveals that recursive refinement improves performance up to a certain number of refinement steps, beyond which error accumulation leads to degradation. This highlights an inherent trade-off in recursive inference and suggests that future work should explore adaptive stopping criteria or theoretically grounded convergence mechanisms for stable refinement.

\clearpage
\bibliographystyle{ACM-Reference-Format}
\bibliography{references}
\clearpage
\onecolumn
\AtBeginDocument{%
  \providecommand\BibTeX{{%
    Bib\TeX}}}

\hypersetup{
    colorlinks=true,
    linkcolor=blue,
    urlcolor=blue,
    citecolor=blue
}

% Tighten document-wide float spacing parameters
\setlength{\textfloatsep}{4pt plus 1pt minus 1pt}
\setlength{\floatsep}{4pt plus 1pt minus 1pt}
\setlength{\intextsep}{4pt plus 1pt minus 1pt}

\appendix
\section{Appendix}
\label{sec:appendix}
This appendix provides ablation results, implementation details, and additional computational analysis for \textsc{RecRec}.

\subsection{Implementation Details}
We implement \textsc{RecRec} in PyTorch with an embedding dimension of 384 for all methods and a maximum sequence length of 50. The outer recursion depth is fixed to $T=7$ and each outer iteration performs $n=3$ inner refinement loops. Following the leave-one-out evaluation protocol, each test instance is paired with 99 randomly sampled negative items, forming a candidate set of 100 items.
Models are trained for 50 epochs using Cross-Entropy loss with the Adam optimizer ($lr=10^{-3}$) and batch size 512. Exponential Moving Average (EMA) with decay 0.999 is used to stabilize optimization. All experiments are conducted on a single NVIDIA RTX 3090 GPU.

% --- ABLATION STUDY SECTION ---
\subsection{Detailed Ablation Study}
Besides the Steam dataset presented in the main paper, we perform ablation studies on the Luxury Beauty and Video Games datasets. We evaluate three variants: (1) No correction gate (G), (2) No inner recursion (R), and (3) No latent variable ($Z$).

\begin{table}[htbp]
\centering
\caption{Ablation Study on Luxury Beauty and Video Games}
\label{tab:ablation_combined}
\vskip 0.05in
\resizebox{\textwidth}{!}{%
\begin{tabular}{l|ccc|ccc|ccc||ccc|ccc|ccc}
\toprule
\multirow{3}{*}{\textbf{Configuration}} & \multicolumn{9}{c||}{\textbf{Luxury Beauty}} & \multicolumn{9}{c}{\textbf{Video Games}} \\
\cmidrule(lr){2-10} \cmidrule(lr){11-19}
& \multicolumn{3}{c|}{HR@K} & \multicolumn{3}{c|}{NDCG@K} & \multicolumn{3}{c||}{Prec@K} & \multicolumn{3}{c|}{HR@K} & \multicolumn{3}{c|}{NDCG@K} & \multicolumn{3}{c}{Prec@K} \\
\cmidrule(lr){2-4} \cmidrule(lr){5-7} \cmidrule(lr){8-10} \cmidrule(lr){11-13} \cmidrule(lr){14-16} \cmidrule(lr){17-19}
& @1 & @5 & @10 & @1 & @5 & @10 & @1 & @5 & @10 & @1 & @5 & @10 & @1 & @5 & @10 & @1 & @5 & @10 \\
\midrule
No G & 0.33 & 0.42 & 0.51 & 0.33 & 0.38 & 0.39 & 0.33 & 0.09 & 0.05 & 0.30 & 0.50 & 0.59 & 0.30 & 0.41 & 0.48 & 0.30 & 0.09 & 0.07 \\
No R & 0.39 & 0.53 & 0.61 & 0.39 & 0.45 & 0.45 & 0.39 & 0.10 & 0.06 & 0.33 & 0.56 & 0.63 & 0.33 & 0.46 & 0.52 & 0.33 & \textbf{0.12} & 0.07 \\
No $Z$ & 0.36 & 0.50 & 0.57 & 0.36 & 0.40 & 0.41 & 0.36 & 0.10 & 0.06 & 0.29 & 0.47 & 0.52 & 0.29 & 0.38 & 0.44 & 0.29 & 0.09 & 0.07 \\
\midrule
\textbf{Full} & \textbf{0.43} & \textbf{0.59} & \textbf{0.66} & \textbf{0.43} & \textbf{0.51} & \textbf{0.53} & \textbf{0.43} & \textbf{0.12} & \textbf{0.06} & \textbf{0.36} & \textbf{0.61} & \textbf{0.71} & \textbf{0.36} & \textbf{0.49} & \textbf{0.52} & \textbf{0.36} & \textbf{0.12} & 0.07 \\
\bottomrule
\end{tabular}
}
\end{table}

\bigskip % Creates a clean, standard separation before the next section
% --- MODEL CONFIDENCE SECTION ---
\subsection{Analysis of Model Confidence}
We define the confidence margin as:
\begin{equation}
s_{t,i}=y_t\mathbf{e}_i^{\top},\qquad M_t=s_{t,\mathrm{target}}-\max_{j\neq\mathrm{target}} s_{t,j}
\end{equation}
where $s_{t,\mathrm{target}}$ denotes the logit of the ground-truth item.

\vspace{12pt} % Explicit space to separate the floating table above from the figure below

\begin{figure}[htbp]
\centering
\includegraphics[width=0.52\linewidth]{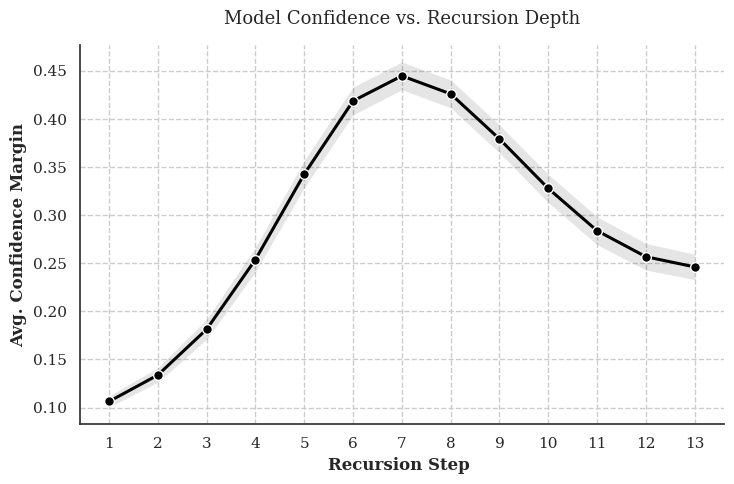}
\caption{Step-wise confidence evolution during recursive refinement.}
\label{fig:confidence_recursion}
\end{figure}

\vspace{12pt}
Figure~\ref{fig:confidence_recursion} illustrates the evolution of the confidence margin during recursive refinement. The confidence increases consistently up to $T=7$, after which additional recursion provides diminishing returns.

\vspace{12pt}
Table~\ref{tab:hr_recursion} evaluates how varying the recursion depth impacts the recommendation performance on the Steam dataset. We observe three key trends from these results:

\begin{table}[htbp]
\centering
\caption{Impact of Recursion Depth on the Steam Dataset}
\label{tab:hr_recursion}
\vskip 0.05in
\begin{tabular}{lc}
\toprule
\textbf{Recursion Steps} & \textbf{HR@1} \\
\midrule
$T=1$ & 0.49 \\
$T=3$ & 0.57 \\
$T=7$ & \textbf{0.59} \\
$T=13$ & 0.55 \\
\bottomrule
\end{tabular}
\end{table}

% --- COMPUTATIONAL EFFICIENCY SECTION ---
\subsection{Computational Efficiency}
Table~6 compares the training time of traditional recommenders, LLM-based recommenders, and \textsc{RecRec}. Our method reduces training cost by more than 98\% compared with existing LLM-based recommendation models while maintaining competitive recommendation accuracy.

\begin{table}[htbp]
\centering
\caption{Training Time Comparison (minutes)}
\label{tab:runtime_full}
\vskip 0.05in
\begin{tabular}{lccc}
\toprule
\textbf{Model} & \textbf{Luxury Beauty} & \textbf{Video Games} & \textbf{Steam} \\
\midrule
SASRec & 40 & 180 & 50 \\
NGCF & 25 & 75 & 40 \\
LightGCN & 35 & 80 & 60 \\
UniRec & 100 & 250 & 180 \\
ReaRec & 90 & 290 & 170 \\
\midrule
Allmrec & 3700 & 12400 & 5600 \\
Tallrec & 6700 & 19800 & 10300 \\
GPT4Rec & 1200 & 5200 & 2300 \\
\midrule
\textbf{RecRec (Ours)} & \textbf{55} & \textbf{210} & \textbf{90} \\
\bottomrule
\end{tabular}
\end{table}

\end{document}